\documentclass[12pt]{iopart}

\usepackage{graphicx}
\usepackage{braket}
\usepackage{cite}
\usepackage{color}
\usepackage{bbm}

\expandafter\let\csname equation*\endcsname\relax
\expandafter\let\csname endequation*\endcsname\relax
\usepackage{amsmath}
\usepackage{iopams}  
\usepackage{layouts}
\usepackage{bbold}

\begin{document}

\title[Quantum processing by remote quantum control]{Quantum processing by remote quantum control}

\author{Xiaogang Qiang$^1$, Xiaoqi Zhou$^{2,1}$, Kanin Aungskunsiri$^1$, Hugo Cable$^1$, Jeremy L. O'Brien$^1$}

\address{$^1$Quantum Engineering Technology Labs, H. H. Wills Physics Laboratory and Department of Electrical \& Electronic Engineering, University of Bristol, BS8 1FD,UK.}
\address{$^2$State Key Laboratory of Optoelectronic Materials and Technologies and School of Physics,
Sun Yat-sen University, Guangzhou 510275, China.}

\ead{qiangxiaogang@gmail.com \\ ~~~~~~~zhouxq8@mail.sysu.edu.cn\\ ~~~~~~~jeremy.obrien@bristol.ac.uk}

\begin{abstract}
Client-server models enable computations to be hosted remotely on quantum servers. 
We present a novel protocol for realizing this task, with practical advantages when using technology feasible in the near term.  Client tasks are realized as linear combinations of operations implemented by the server, where the linear coefficients are hidden from the server. 
We report on an experimental demonstration of our protocol using linear optics, which realizes linear combination of two single-qubit operations {by a remote single-qubit control}. {In addition, we explain when our protocol can remain efficient for larger computations, as well as some ways in which privacy can be maintained using our protocol.}
\end{abstract}

\section{Introduction}
\noindent Quantum computing offers the possibility of achieving substantial algorithm speedups compared to classical computing~\cite{shor1997polynomial,grover1997quantum,Montanaro2016}, and can preserve the privacy of computations while doing so. Given the intrinsic difficulties in building a quantum computer, this privacy preservation will be crucial for any client-server model, which will likely provide a practical and efficient way to access quantum computing resources. In the scenario where a client delegates his computation to a quantum server, the data can readily be hidden from the server by using algorithms designed to work on encrypted data~\cite{Fisher2014,aharonov2008interactive,childs2005secure,dupuis2012actively,broadbent2013quantum}. A protocol for ``blind'' quantum computing, based on the paradigm of measurement-based quantum computing~\cite{arrighi2006blind,broadbent2009universal}, was recently demonstrated using linear optics~\cite{barz2012demonstration}.  Here the client implements an algorithm by requesting that the server performs consecutive adaptive single-qubit measurements on a (large) blind cluster state---a multi-particle entangled state created from qubits transmitted by the client. Since the states of the transmitted qubits are chosen randomly by the client, the computations on the blind cluster state do not reveal any data or the algorithm to the server~\cite{barz2012demonstration}. The randomness source that is used by the client should be carefully examined to avoid any correlations with the server and must achieve high-speed operation (such as was recently reported in ref~\cite{abellan2016quantum}). Full-scale demonstrations of this blind quantum computing protocol would also require that the server has the ability to create large cluster states, which is beyond the capabilities of current quantum technologies.

Here we propose a fundamentally new type of protocol for allowing clients to execute quantum processing on {a remote} server. In our approach, the client translates his task into a linear combination of quantum operations performed by server. Arbitrary unitary operations can be represented in a linear-combination form using the Cartan decomposition~\cite{kraus2001optimal}.   The linear coefficients are then encoded in a quantum state, and transmitted from client to server using quantum teleportation. {\ As we will argue, the client can keep the linear coefficients hidden from the server.}  
To enable the required linear combining of quantum operations in our protocol, we will utilise circuits based on a technique to add coherent control to arbitrary (unknown) quantum operations, demonstrated in Ref.~\cite{zhou2011adding}.  This technique is based on gates which can exploit extensions of the logical Hilbert space used for computation.  We will proceed as follows: we will first explain circuits for realising linear-combinations of a fixed family of quantum operations, before explaining in detail how they can be used to enable quantum computation in a client-server model.  Then we will report a proof-of-principle experimental demonstration of our protocol in a linear-optic setup, which implements arbitrary linear combinations of two single-qubit quantum operations by a remote one-qubit control.

\section{Linear combining of quantum operations}

\begin{figure}[tbp]
\centering
\includegraphics[scale = 0.75]{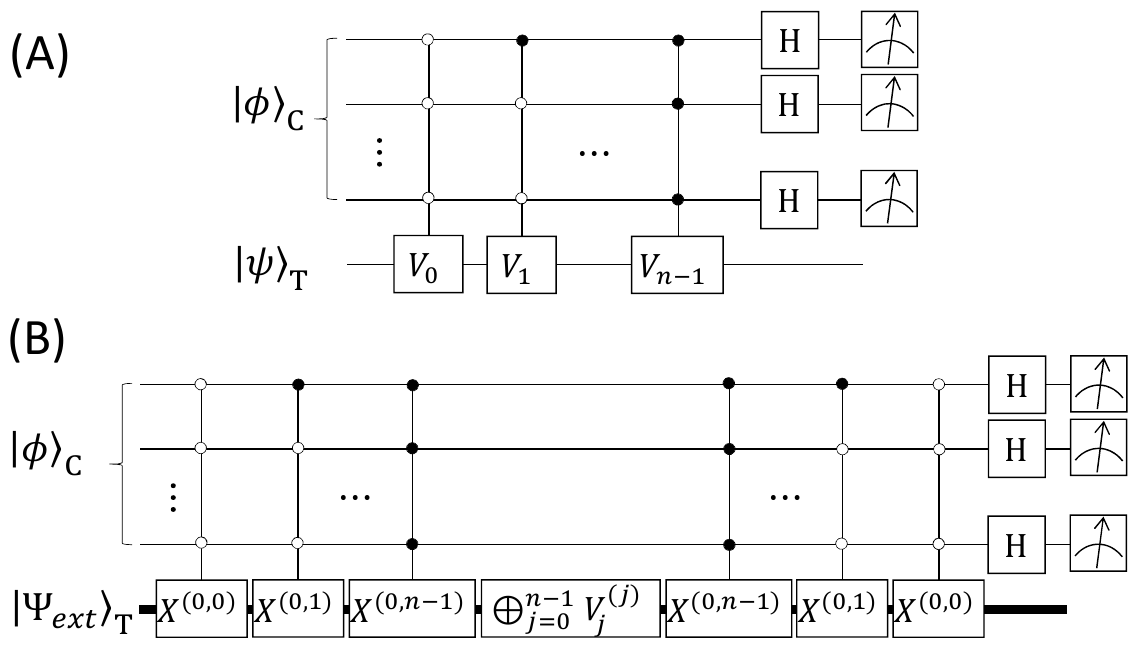}
\caption{
\textbf{Implementing linear-combination operations:} (A) Circuit for implementing linear-combination operations which assumes the availability of multiply-controlled $V_j$ gates. There are $k$ control qubits with initial state $\ket{\phi}_C = {\sum\nolimits_{j = 0}^{n - 1} {{\alpha _j}\ket{j}_C} }$ and $n=2^k$. $T$ is a $d$-dimensional target system. $U_T = {\sum\nolimits_{j = 0}^{n - 1} {{\alpha _j}V_j} }$ acts on $T$ when the measurement outcome is $\ket{0}\!\bra{0}_C^{\otimes k}$. (B) The LCC implements the same conditional operation as in (A) but without controlled $V_j$ gates, with $T$ extended to $(n\times d)$-dimensions, using operations on subspaces of $T$.
}
\label{QcircuitLCU}
\end{figure}

Suppose that we want to implement some unitary $U_T$ which can be expressed in the form,
\begin{align}
U_T = \sum\nolimits_{j = 0}^{n - 1} {{\alpha _j}V_j} ,
\end{align}
where the $V_j$ are gates acting on a $d$-dimensional target ($T$) subspace, and the $\alpha_j$ are complex coefficients satisfying
\begin{align}
{\sum\nolimits_{j = 0}^{n - 1} {|{\alpha _j}|^2} }=1.
\end{align}
When controlled-$V_j$ gates are available, we can implement $U_T$ probabilistically through the circuit illustrated in Fig.~\ref{QcircuitLCU}(A). Here the $\alpha_j$ are encoded in the initial state for the $k$-qubit control (C),
\begin{align}
\ket{\phi}_C = {\sum\nolimits_{j = 0}^{n - 1} {{\alpha _j}\ket{j}_C} }, \label{controlState}
\end{align}
where $n= 2^k$ and $j$ labels the computational basis, and the circuit succeeds when all control qubits are measured to be 0 in the computational basis at the end.

However, this approach for implementing $U_T$ cannot work when the $V_j$'s must be assumed to be black-box operations, due to a no-go theorem which states that adding control to unknown quantum operations is impossible in the (conventional) quantum circuit model~\cite{araujo2014quantum,thompson2013quantum}: any protocol which attempts to add control to a black-box operation must be able to differentiate $V_j$ and $\exp(i\theta)V_j$, but standard quantum circuits always generate identical measurement outcomes for these two cases. Nonetheless,  control can be added in many systems, by exploiting the fact that physical operations often act non-trivially on some degrees of freedom or subspaces of quantum states, while acting trivially on others. The description of $V_j$ for such cases should be modified to $V_j \oplus I$, and control can be added even when this extension is one dimensional~\cite{araujo2014quantum}. It has been shown that control qubits can be simply added to a single-qubit unitary by moving part of the state of a target qubit into an expanded Hilbert space \cite{lanyon2009simplifying}. A more general scheme was proposed in reference \cite{zhou2011adding} for adding control to an arbitrary quantum operation, with the implementation of its optical version based on the controlled-path (CP) gate~\cite{lin2009quantum} that controls the target photon's path conditioned on the control photon's polarization. The CP gate was first proposed for realizing quantum controlled gates in the context of weak optical cross-Kerr nonlinearities~\cite{lin2009single,lin2009processing}. Techniques based on expanding the computational Hilbert space have also been demonstrated for adding control for subroutines of quantum computation~\cite{zhou2013calculating} and implementing the Fredkin gate~\cite{patel2016quantum}. Here we use the same techniques to implement a linear-combination circuit (LCC) which is illustrated in Fig.~\ref{QcircuitLCU}(B).

LCCs can exploit black box unitaries to implement a target quantum evolution using coherent control, using the control state as in Eq.(\ref{controlState}), acting on a $(n\times d)$-dimensional target subspace $T$. $T$ decomposes into $n$ $d$-dimensional subspaces, with the $j^{th}$ subspace is spanned by basis elements $\{ \ket{jd}_T, \cdots, \ket{(j+1)d-1}_T\}$. The LCC uses a series of subspace-swap operations, $X^{(0,j)}$ (which exchange corresponding basis elements for the $0^{th}$ and $j^{th}$ subspaces) which are controlled by qubits in $C$, and performs the sum operation $\oplus _{j = 0}^{n - 1}{V_j^{(j)}}$, where $V_j^{(j)}$ implements the same operation as $V_j$ previously but on the $j^{th}$ subspace of $T$. The initial state for $T$ is taken to be
\begin{align}
\left| \Psi_{ext}  \right\rangle_T = {\sum\nolimits_{j = 0}^{d - 1} {{\beta _j}{\left| j \right\rangle_T}} } + {\sum\nolimits_{j = d}^{nd - 1} {{0}{\left| j \right\rangle_T}} }.
\end{align}
Following the step-by-step evolution given in Supplementary Material, it is straightforward to verify that, when the control qubits are all measured to be 0 in the computational basis, the target evolves according to:
\begin{align}
\left| \Psi_{ext}  \right\rangle_T \to \sum {\alpha_j V_j^{(0)}\ket{\Psi_{ext}}_T}.
\end{align}
Note here $V_j^{(0)}$ implements $V_j$ on the $0^{th}$ subspace of $T$ as defined before. The success probability is readily found to be $1/n$, which is independent of the size of the $V_j$.

Any arbitrary quantum unitary operation can in principle be decomposed into a linear sum of elementary operations. Using Cartan's KAK decomposition, we can explicitly rewrite any two-qubit unitary operation, $U_{\text{SU(4)}}$, as a linear combination of four tensor products of two single-qubit gates. { Furthermore, Cartan's decomposition allows an $n$-qubit unitary operation $U_{\text{SU}(2^n)}$ to be recast as a linear combination of tensor products of $n$ single-qubit gates~\cite{khaneja2001cartan}. Such a decomposition is, in general, not efficient, in the sense that there may be exponentially-many terms.} And thus, the success probability of LCC for general $U_{\text{SU}(2^n)}$ can be exponentially small. { However, for some { non-trivial families of} unitary operations the linear decomposition method can be efficient. For example, an $n$-qubit controlled-unitary gate CU can be decomposed as $\frac{I+\sigma_z}{2}\otimes I + \frac{I-\sigma_z}{2} \otimes U$ where $U$ is an $(n-1)$-qubit operation~\cite{zhou2011adding}. Only one control qubit is required to implement this operation and high success probability can be obtained. { Although the number of linear-combining terms is restricted, the size of each term can be large and reconfigurable, providing sufficient computing power and flexibility for various applications.}} 
{ It is worth noting that the proposed LCC can also be interpreted by using the notion of duality quantum computation~\cite{gui2006general,long2007mathematical,wei2016duality}, which was originally proposed to exploit the wave-particle duality and then developed to work within the framework of conventional quantum computing.}

\section{Implementing quantum processing by remote quantum state control}
\begin{figure}[tbp]
\centering
\includegraphics[scale = 0.55]{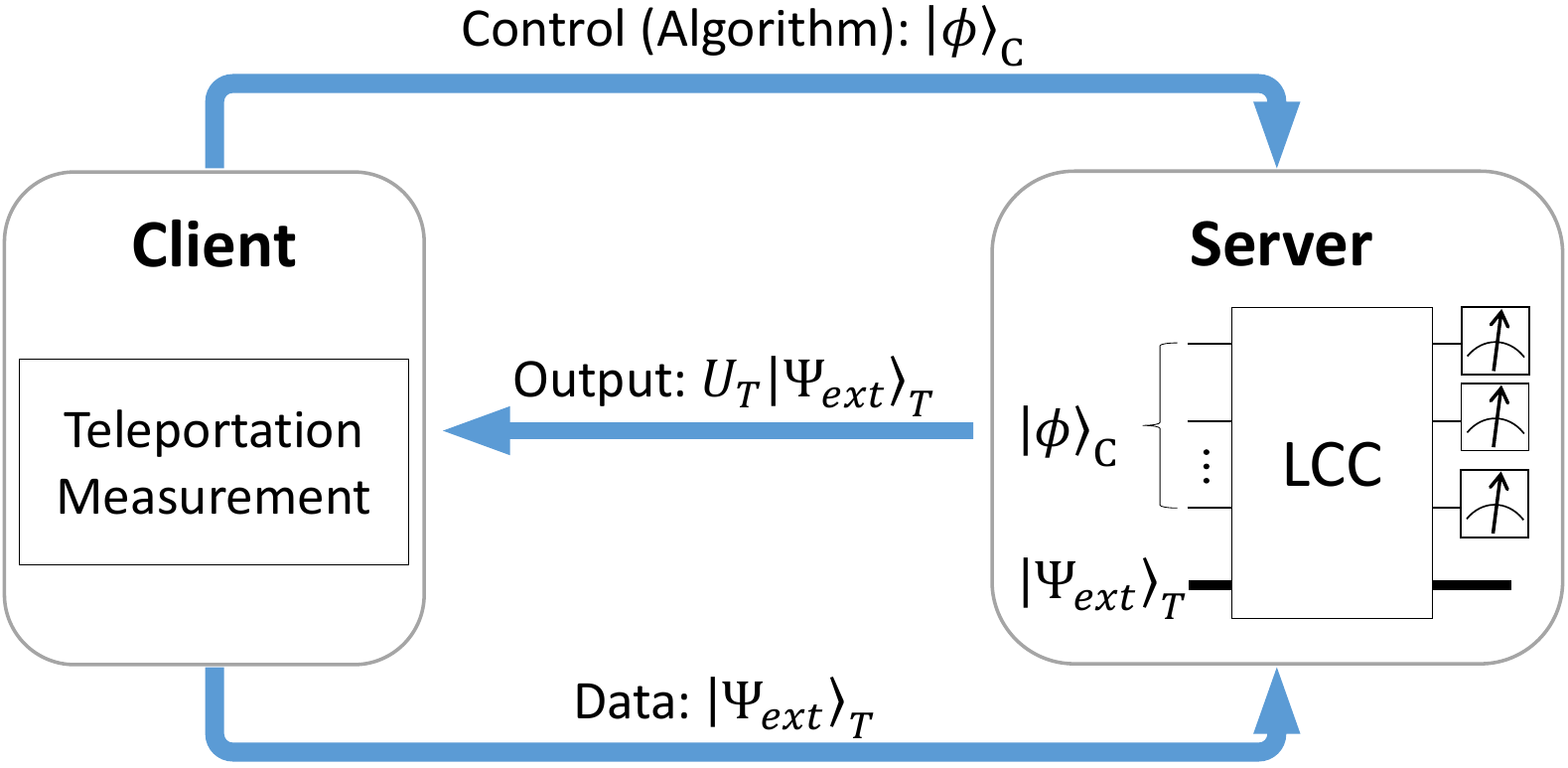}
\caption{
\textbf{Protocol for remote quantum processing:} For each of the client's requests, the server first repeatedly runs the LCC till it succeeds. The client then teleports a quantum control state $\ket{\phi}_C$ to the server using quantum teleportation EPR channels to complete his computation. The computational input $\ket{\Psi_{ext}}_T$ can be transmitted to the server (and the computational output $U_T\ket{\Psi_{ext}}_T$ back to the client) using additional quantum-teleportation channels or direct transmissions.
\label{scheme}}
\end{figure}

The LCC described above provides a way to implement quantum information processing using a client-server model, as illustrated in Fig.~\ref{scheme}. We assume now the $V_j$'s are the computational resources provided by the server and the $\alpha_j$'s are  configured by the client to encode an algorithm. The $\alpha_j$'s are encoded into the control state $\ket{\phi}_C$ and transmitted from the client to the server remotely.
The transmission of states between the client and the server is performed by a (multi-)qubit teleportation protocol \cite{bennett1993teleporting,chen2006general} using generalised Bell measurements.  The control state $\vert \phi\rangle_{\rm C}$ has $k$ qubits, and $k$ EPR channels must be shared between the client and server to enable teleportation of this state.  Similarly, $\lceil \log_2 d \rceil$ EPR channels are required to teleport the computational input $\ket{\Psi_{ext}}_T$ from client to server, and a further $\lceil \log_2 d \rceil$ EPR channels are required to teleport the computational output from server to client ($d$  is defined as previously).  To start the computation, the client requests the server to run the LCC, and the server repeatedly runs the LCC on the EPR channels (resetting them as required). When the LCC succeeds, the server informs the client and performs teleportation measurements on the LCC output and corresponding EPR channels.  Finally, the client performs teleportation measurements on $\ket{\phi}_C$ and $\ket{\Psi_{ext}}_T$ (and the corresponding EPR channels).  When all LCC and teleportation steps succeed, $U_T \ket{\Psi_{ext}}_T$ is returned to the client.

\begin{figure}[tbp]
\centering
\includegraphics[scale =0.9]{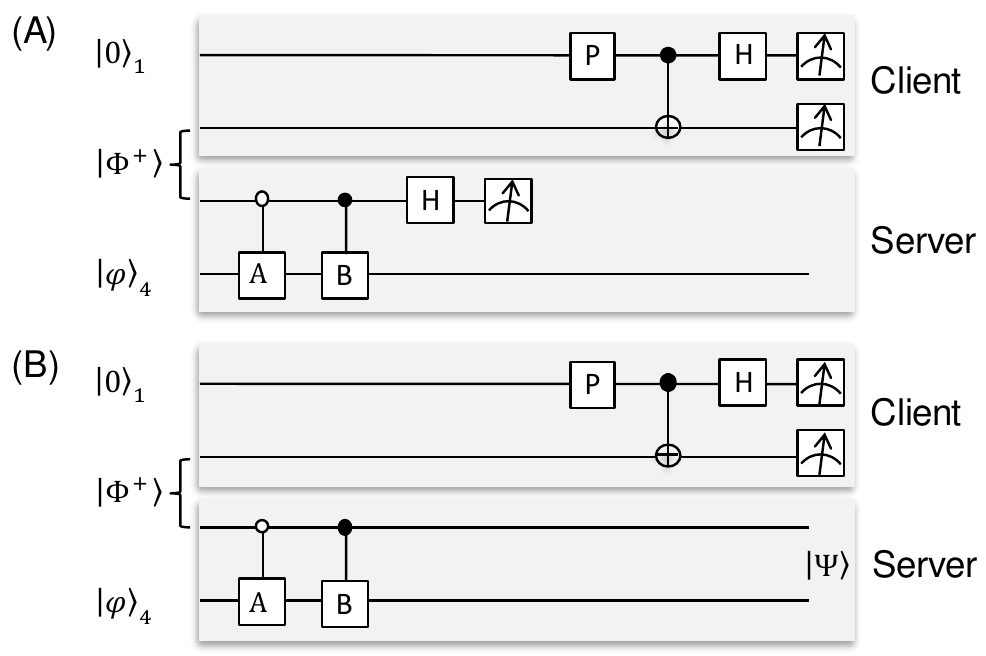}
\caption{
\textbf{Security analysis for one-qubit control quantum processing.} (A) $\ket{\Phi^+} = \frac{1}{\sqrt{2}}(\ket{0_2}\ket{0_3}+\ket{1_2}\ket{1_3})$ is the EPR state shared between the client and the server. $A$ and $B$ are two arbitrarily-large quantum operation of the same size. $\ket{\varphi}_4$ is the input state for the client's computation. We ignore the teleportation process of $\ket{\varphi}_4$ from the client to the server. $P$ is a local single-qubit operation to configure the one-qubit control state $\ket{\phi}_C$. The server repeatedly runs the LCC until he measures the local control qubit (qubit 3) to be ``0'', and then he informs the client to start the configuration and teleportation of the control state. 
(B) In this case, the server tries to cheat by not performing the measurement on the local control qubit, and directs the client to start the teleportation process. $\ket{\Psi}$ represents the state of remaining qubits that the server obtains when the quantum teleportation succeeds. A step-by-step evolution is shown in Supplementary Material.
}
\label{fig:security}
\end{figure}

{ By keeping the control state $\ket{\phi}_C$ hidden from the server, this protocol can provide security for the client's computation.} We first consider the simplest case where the client only sends a one-qubit control state to the server so that a linear combination of two quantum operations $A$ and $B$ can be implemented. The corresponding quantum circuit is shown in Fig.~\ref{fig:security}(A), where we assume that $A$ and $B$ are not black-box operations and also ignore the teleportation of the input state for the computation. The circuit starts from the initial state $ \frac{1}{\sqrt{2}}\ket{0}_1(\ket{0}_2\ket{0}_3+ \ket{1}_2\ket{1}_3)\ket{\varphi}_4$. In the case where the server follows the protocol, the server first runs the LCC until it succeeds---the qubit 3 (local control qubit) is then measured to be ``0'' in computational basis. The state of remaining qubits is $\frac{1}{\sqrt{2}}(\ket{0}_1\ket{0}_2A\ket{\varphi}_4+\ket{0}_1\ket{1}_2B\ket{\varphi}_4)$. The client then performs the quantum teleportation. When he measures the qubit 1 and qubit 2 to be ``0'' in computational basis, the state of remaining qubit becomes $(\alpha A + \beta B)\ket{\phi}_4$ immediately. During the whole process, the server does not have any chance to detect the control state (encoded in the qubit 1 by the client's local operation $P$), because he needs to measure the local control qubit (qubit 3) before the client performs the configuration of control.

{ Next we consider the case where the server does not perform the measurement on the local qubit before the teleportation as our protocol demands.}
In this case, the circuit will evolve as shown in Fig.~\ref{fig:security}(B). When the client measures the qubit 1 and qubit 2 to be ``0'', the state of remaining qubits will be $\alpha \ket{0}_3A\ket{\varphi}_4 + \beta \ket{1}_3 B \ket{\varphi}_4$ (we denoted it as $\ket{\Psi}$). Now the question is that whether the server can extract the information of the control state $\ket{\phi}_C = \alpha \ket{0} + \beta \ket{1}$ without being detectable to the client. To achieve this, the server needs to extract $\ket{\phi}_C$ and also output the correct result of the computation $(\alpha A + \beta B)\ket{\varphi}_4$ to the client. In other words, the server needs to find an operation $U_s$ satisfying
\begin{align}
(\alpha \ket{0}+ \beta \ket{1})(\alpha A + \beta B)\ket{\varphi} = U_s (\alpha \ket{0}A\ket{\varphi} + \beta \ket{1} B \ket{\varphi}).
\end{align}  
Such an operation $U_s$ does not exist for unknown parameters $\alpha$ and $\beta$, because it would allow copying of an unknown quantum state which violates the no-cloning theorem~\cite{wootters1982single,dieks1982communication}. 
{ However, it is possible for the server (or a third party) to generate a copy of the control state with imperfect fidelity, for example, by using a universal quantum cloning machine (UQCM)~\cite{buvzek1996quantum,gisin1997optimal} even with a single copy of the control state. Such cloning attacks are difficult to prevent since they could be disguised as channel loss, and thus can lead to leaking of information about the client's computation.}

{
{ 
For many applications such as Shor's factorization algorithm~\cite{shor1997polynomial} and Grover's search algorithm~\cite{grover1997quantum}, the client can get the result by just running the protocol a few times. Then the server (or a third party) might potentially obtain partial information about the control state by using UQCM. 
For applications that require many runs of the protocol, the client would need to send excess copies of the control state, and thus the server might potentially gain complete information about the control state, for example, by using quantum state tomography. To address this vulnerability we present a modified protocol below:}

For a computation with the control state $\rho = \ket{\phi}_C \bra{\phi}_C$, define a decoy state
\begin{align}
\rho_m=\frac{1+\epsilon}{n}\mathbb{1} - \epsilon \rho 
\end{align}
where $n$ is the number of dimensions of $\rho$ and $0< \epsilon \le 1/(n-1)$. $\rho_m$ can be generated by sending its eigenstates with probabilities given by corresponding eigenvalues. On each run of the protocol, the client sends the control state $\rho$ with probability $\epsilon/(1+\epsilon)$ and {the decoy state} $\rho_m$ with probability $1/(1+\epsilon)$. As the client knows exactly what state he sent each run, he can just discard the output states corresponding to the decoy states and keep the correct ones for further applications. From the perspective of the server, the state received will be   
\begin{align}
\frac{\epsilon}{1+\epsilon}\rho + \frac{1}{1+\epsilon}\rho_m = \frac{1}{n}\mathbb{1}. 
\end{align}
The state $\mathbb{1}/n$ has the maximal entropy (= $\log{n}$), implying that the server has no knowledge about the received states at all.}

{
The client can verify the result directly for certain applications (e.g. Shor's factorization and Grover's search) but not others (e.g. some large quantum simulations). However, the client is still able to verify (or monitor) the computation process for applications whose results cannot be verified directly. We have shown that the decomposed component $V_i$ can be as simple as a tensor product of single-qubit gates and can therefore be verified with limited resources. Throughout the full computation process, the client can randomly send each basis state $\ket{i}$ ($i=0,1,\cdots,n-1$) to the server, and since only the corresponding component $V_i$ is applied, the output can be checked (via state tomography or measurements in multiple bases). This approach allows the client to diagnose whether the server is running the LCC correctly, and it can be combined with the strategy above for preventing the control state from being measured by the server (or a third party): the client chooses a proportion of the runs of the protocol for performing computation and the rest of the runs of the protocol for verification. Assuming the proportion of runs of the protocol for computation to be $\tau$ ($0<\tau < 1$), the client would send the control state $\rho$ with probability $\tau \epsilon /(1+\epsilon)$, the decoy state $\rho_m$ with probability $\tau/(1+\epsilon)$, and each basis state $\ket{i}$ with probability $(1-\tau)/n$ on each run. The state the server receives is then
\begin{align}
\tau\left(\frac{\epsilon}{1+\epsilon}\rho + \frac{1}{1+\epsilon}\rho_m\right)+ \frac{1-\tau}{n} \sum_{i=0}^{n-1}{\ket{i}\bra{i}} = \frac{1}{n}\mathbb{1}. \label{eq:mix_verify}
\end{align}
{ Therefore, although the whole computation process takes longer, the server is given no information about whether the states it receives are for verification purposes or for performing an algorithm, and no information about the control state.} If the server intercepts a fixed proportion of the control qubits in a way which randomizes the results, the probability that the server is not detected is suppressed exponentially as the number of runs of the protocol grows.}

We have shown that the success probability of the LCC decreases exponentially with the number of control qubits. However, in the secure quantum processing protocol, the server only needs to inform the client when the LCC succeeds, ensuring that the LCC works with $100\%$ success probability from the standpoint of the client. The success probability for teleporting the control state exponentially decreases with the number of teleported qubits, implying poor scaling with large control states. Therefore, {our protocol is practical only for small control states, i.e. the number of linear terms $n$ should be polynomial-sized with respect to the problem size. For a typical case of the modified protocol combining verification and computation where $\epsilon = 1/(n-1)$ and $\tau = 1/2$, the probability of the client sending the control state $\rho$ for each run will be $1/2n$, and thus the number of runs of the protocol required will be $O(2n)$ times more than the original protocol, which brings only polynomially-increasing cost}. { The whole client-server computation scheme could (where required) include the quantum teleportation of the computation input and output. Teleporting the output has 100\% success probability with necessary correction operations, while the success probability of teleporting the input depends on the dimension $d$ of the target operation (specifically, equals to $1/d^2$) since the correction operations generally do not commute with the target operation. Taking these teleportation steps into account, the success probability of the whole scheme is $1/O(\mbox{poly}(nd))$.
The client here is required to have the capability to create small control states, which is trivial compared to the capabilities that the server must have. It is also noteworthy that the success probability could be further improved by using port-based teleportation (rather than conventional quantum teleportation)~\cite{ishizaka2008asymptotic,ishizaka2009quantum}, which transmits a one-qubit state to one of $K$ output ports using $K$ EPR pairs and is asymptotically faithful and deterministic for large $K$. 
}

\section{Experimental demonstration}
\begin{figure*}[tbp]
\centering
\includegraphics[width=1\textwidth]{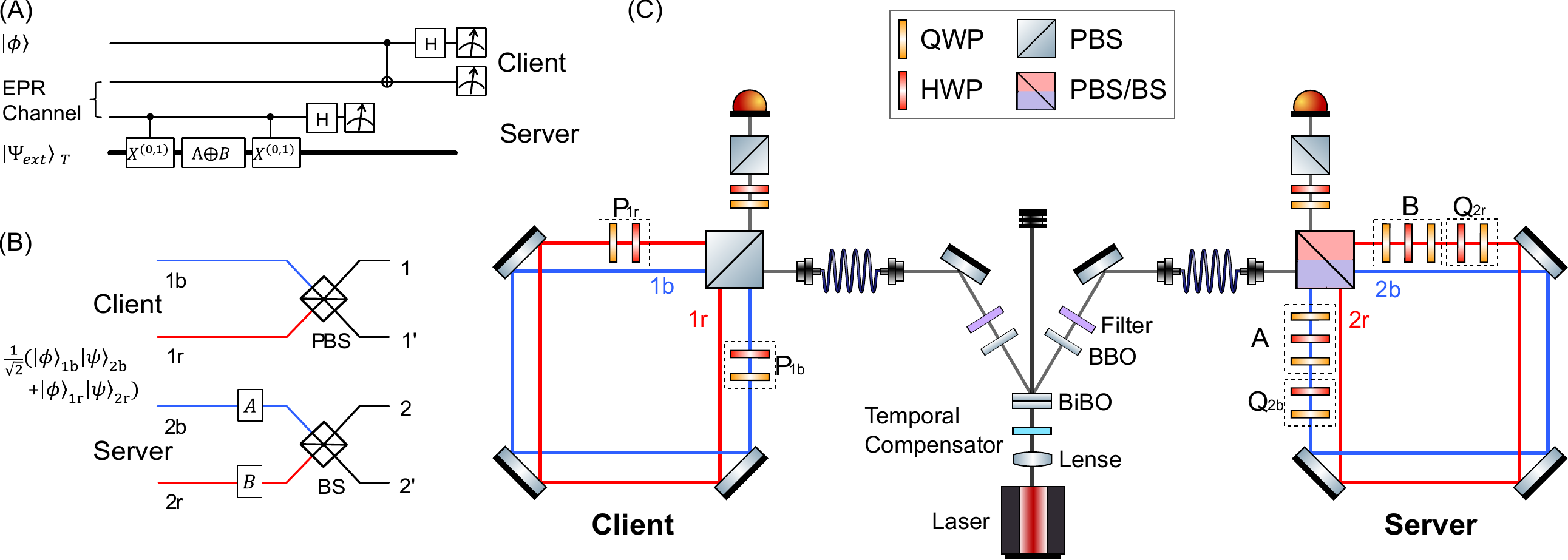}
\caption{
{\textbf{Experimental demonstration:}
(A) Circuit for implementing quantum processing by remote one-qubit quantum control.
(B) Schematic for optical implementation of (A). Client and server share a pair of spatially-entangled photons: $( \left| {\phi_{1b}\psi_{2b}} \right\rangle + \left| {\phi_{1r}\psi_{2r}} \right\rangle)/{\sqrt 2}$. When the photons exit at port 1 and 2, the output state of the photon on server's side will be $(\alpha A + \beta B) \left| \psi \right\rangle$, where $\alpha$ and $\beta$ are controlled by client's one-qubit control state $\ket{\phi} = \alpha \ket{0}+\beta \ket{1}$.
(C) In our setup, entangled photon pairs are generated by a SPDC source using paired type-I BiBO crystal in a sandwich configuration. $P_{1b}$ and $P_{1r}$ ($Q_{2b}$ and $Q_{2r}$) configure $\ket{\phi}$ ($\ket{\psi}$). $A$ and $B$ can implement arbitrary single-qubit gates. Further details are given in Appendix.
}}
\label{setup}
\end{figure*}

Here we report on a demonstration of our protocol using a linear-optic setup, which realises a circuit for generating linear combinations of two single-qubit gates with one-qubit quantum control, as shown in Fig.~\ref{setup}(A).  Our experimental setup exploits both path and polarization degrees of freedom of photons. Since direct implementation of controlled-$V_j$'s is very challenging using current technology, we demonstrate a LCC using the method shown in Fig.~\ref{setup}(B).  To understand how it works, suppose that server starts with a single photon in the state
\begin{align}
\alpha \left| \psi \right\rangle_{b} \left| \mbox{vac} \right\rangle_r + \beta \left| \mbox{vac} \right\rangle_b \left| {\psi} \right\rangle_r , \label{serverPhotonState}
\end{align}
where $\left| \psi \right\rangle$ is an (arbitrary) polarization-encoded qubit, $b$ and $r$ label the blue and red spatial modes, and $\left| \mbox{vac} \right\rangle$ represents unoccupied modes (and will be dropped below). Two single-qubit gates $A$ and $B$ act only on photon in the blue or red path respectively, yielding the state: $\alpha A \left| \psi \right\rangle_{b} + \beta B \left| {\psi} \right\rangle_r$.
The blue and red modes are then mixed on a (non-polarising) beam splitter (BS) to remove path information. In the case where the photon exits at port 2, the output state of the photon which is obtained is $(\alpha A + \beta B) \left| {\psi} \right\rangle$, which corresponds to the action of linear combination $\alpha A + \beta B$ on $\vert\psi\rangle$.

{In the remote quantum processing scenario,} client and server start by sharing a pair of entangled photons in state
\begin{align}
\big( \left| {\phi} \right\rangle_{1b}\left| {\psi} \right\rangle_{2b} + \left| {\phi} \right\rangle_{1r}\left| {\psi} \right\rangle_{2r}\big)/{\sqrt 2},
\end{align}
where $\left| \phi \right\rangle = \alpha \left| H \right\rangle + \beta \left| V \right\rangle$ (client photon) and $\left| \psi \right\rangle$ (server photon) encodes a qubit in the polarization basis.  When the blue and red modes of client's photon are mixed on a polarising beam splitter (PBS), the client-server state becomes
\begin{align}
\left|D \right\rangle_{1} (\alpha \left| {\psi} \right\rangle_{2b} + \beta \left| {\psi} \right\rangle_{2r} )+ \left| D \right\rangle_{1'}(\alpha \left| {\psi} \right\rangle_{2r} + \beta \left| {\psi} \right\rangle_{2b}),
\end{align}
where $\ket{D}=(\ket{H}+\ket{V})/{\sqrt 2}$, and contributions corresponding to anti-diagonal polarization at $1$ and $1^\prime$ have been dropped (corresponding to postselection on detection outcomes with diagonal-polarization only).  In the case where client's photon exits at port 1, the state of the server's photon is given by Eq.~\eqref{serverPhotonState}, and the operation $\alpha A+\beta B$ is implemented as above. The experimental setup is shown in Fig.~\ref{setup}(C), and the details are shown in Appendix.

\begin{figure*}[tbp]
\includegraphics[width=1\textwidth]{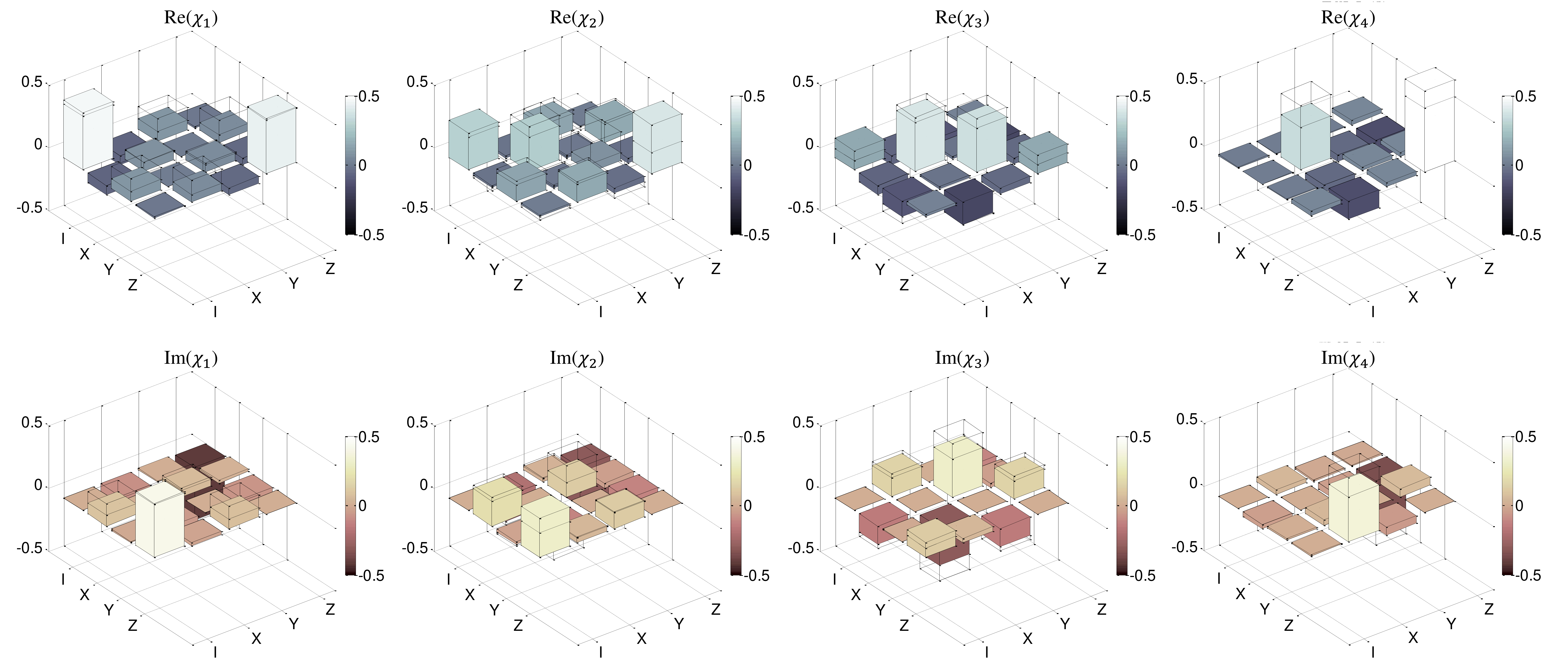}
\caption{
{\textbf{Experimental reconstructed $\chi$ matrices with corresponding theoretical predictions overlaid:}
Three unitary operations
$U_1 = 0.9239A + 0.3827B$,
$U_2 = 0.7071A + 0.7071B$,
$U_3 = -0.3827A+ 0.9239B$
and one non-unitary operation
$U_4 = 0.7071X + 0.7071iZ$ were tested. The corresponding process matrices $\chi_1$, $\chi_2$, $\chi_3$ and $\chi_4$ are shown with their theoretical values overlaid.  We observed process fidelities $94.38\pm0.87\%$, $94.79\pm 0.85\%$, $95.98\pm0.73\%$ and $88.56\pm1.58\%$ respectively. The errors are estimated by adding random noise to the raw data and performing many reconstructions. Further results are given in Supplementary Material.
}}
\label{results1-4}
\end{figure*}

It is worth noting that an arbitrary single-qubit quantum operation $U_{\text{SU(2)}}$ can be implemented as
\begin{align}
U_{\text{SU(2)}} = \alpha_0 I + \alpha_1 \sigma_x + \alpha_2 \sigma_y + \alpha_3 \sigma_z
\end{align}
where $\sigma_x$, $\sigma_y$ and $\sigma_z$ are Pauli matrices, and $\alpha_i$ are complex coefficients satisfying $\sum_{i=0}^{3} |\alpha_i|^2 = 1$ (see details in Supplementary Material). Therefore, linear combination of four gates would be required to implement an arbitrary single-qubit operation if the server were to provide only Pauli gates as the resource to the client. In our experimental setup, the two single-qubit gates provided by the server can be arbitrarily configured, which allows us to demonstrate the secure realization of a wide range of linear-combination operations. We tested a series of linear-combination operations where the two single-qubit gates are set to be
\begin{align}
A = \left( {\begin{array}{*{20}{c}}
\frac{1-i}{{\sqrt 2 }}&{ 0}\\
{ 0}&\frac{-1-i}{{\sqrt 2 }}
\end{array}} \right),
B = \left( {\begin{array}{*{20}{c}}
{0}&\frac{1+i}{{\sqrt 2 }}\\
\frac{1-i}{{\sqrt 2 }}&{0}
\end{array}} \right).
\end{align}
The linear combinations of $A$ and $B$ are always unitary when the client's one-qubit control state has real amplitudes. Our main results are shown in Fig.~\ref{results1-4}, and additional results are also given in Supplementary Material.
Our protocol also allows the client to implement non-unitary operations (even though the server provides only unitary gates). For example, when the two gates $A$ and $B$ are set to be $X$ (Pauli-X) and $Z$ (Pauli-Z) gates respectively, the client can implement non-unitary operation $(X+iZ)/{\sqrt 2}$ by teleporting one-qubit quantum control $\ket{\phi}_C=(\ket{0}+i\ket{1})/{\sqrt 2}$.
To evaluate the performance of each the operations we tested, we performed quantum process tomography and reconstructed corresponding process ($\chi$) matrices from the experimental data, using the maximum-likelihood-estimation technique.
As shown in Fig.~\ref{results1-4}, all of the reconstructed process matrices achieve high process fidelities compared to the corresponding ideal cases.

{

Our experiment serves as a proof-of-principle demonstration of the essential part of our protocol---a remote control state can be used to implement the linear-combining operation. As we mentioned above, the server (or a third party) could use a UQCM to extract partial information about the control state. Also, as post-selection was used in the experiments to choose cases where the teleportation of the control state and the LCC succeed simultaneously, the server can obtain extra copies of the control state by disguising his measurements as failures of the LCC, leading to potential information leak of the control state.

The proposed modified protocol aims to eliminate possible leak of the control state, but requires enhanced capability of the experimental setup. In particular, it costs much increased experimental time to generate the required mixed states and thus needs improved robustness and stability---which would be challenging for our current bulk-optical setup (but could potentially be achieved in a future experiment using integrated photonic waveguide techniques~\cite{politi2008silica,carolan2015universal,wang2016chip}). Possible issues for future demonstration of the modified protocol include experimental imperfections, loss in transmission channels and the photon source. Imperfections in the server's gates (such as $A$, $B$, $Q_{2b}$, $Q_{2r}$ shown in Fig.~\ref{setup}(C)) do not affect the security of the protocol, rather just the outcome of the computation. Imperfections in the client's gates (such as $P_{1b}$, $P_{1r}$ shown in Fig.~\ref{setup}(C)) can affect the creation of the mixed state $\mathbb{1}/n$ (and also potentially mimic effects of a malicious third party or server) and thereby reduce the security offered by the modified protocol.
However, loss in the transmission channels would not cause any added security issue for the modified protocol, since it would just act as a normalization factor for the mixed state $\mathbb{1}/n$. The SPDC photon source creates photon pairs probabilistically, which can be viewed as being equivalent to loss in the channels from a deterministic source, and the security is similarly unaffected by this. A completely quantitative security analysis is beyond the scope of this work and is for future research.

}

\section{Conclusion}

In summary, we have described and demonstrated a novel protocol, which can enable a client to implement complex quantum processing on a remote server without revealing the precise algorithm to the server.
{We leave as an interesting open question whether unconditional security can always be guaranteed using our protocol, which will require an information-theoretic analysis of diverse attacks on the security, as well as the effects of experimental imperfections, such as multi-pair contributions to the state generated by the SPDC source. }Although our discussion has focused on protecting the privacy of the client's algorithm, it can be extended to protect the privacy of the client's data by exploiting existing encryption schemes \cite{Fisher2014}. {Our protocol cannot always achieve efficient implementation of arbitrary quantum circuits (efficient universality), but it could be suitable for some practicable applications, for example, adding control to a remote operation, with less resources and experimental difficulties.}
The LCC circuits used by our protocol are based on decompositions into linear combinations of elementary gates, and differ greatly from the circuits generated by the Solovay-Kitaev algorithm \cite{Dawson2005Solovay-Kitaev} for example. Compared with more conventional techniques to implement quantum computation, such linear-combination-based methods would lead to greater efficiency for some problems: Several works have shown that simulations of Hamiltonian dynamics based on linear combinations of unitary operations can achieve exponentially-improved precision-dependence compared to the conventional product-formula-based algorithms~\cite{Childs2012Hamiltonian,kothari2014efficient}, and even nearly-optimal dependence on all parameters~\cite{berry2015hamiltonian}. By using the linear-combination technique, the dependence on precision can be exponentially improved~\cite{childs2015quantum} compared to the Harrow-Hassidim-Lloyd algorithm~\cite{harrow2009quantum} for the quantum linear systems problem. It can also reduce the query complexity and improve precision for simulations of open quantum systems~\cite{wei2016duality} based on linear combinations of Kraus operators~\cite{nielsen2010quantum}. These applications generally require linear combinations of a great number of unitary operations. It is an interesting open question whether there exist some particular instances that can critically benefit using only a limited number of linear terms. { Considering the alternative interpretation of the LCCs in duality quantum computation, our protocol could be treated as an interesting and important application of duality quantum computation. }Finally, the protocol we have demonstrated here can be implemented in a wide range of physical systems. For example, future photonic demonstrations of our protocol could exploit time-bin and orbital angular momentum degrees of freedom (which can offer high-dimensional quantum subspaces) to implement complex controlled operations.

\section*{Acknowledgements}
The authors would like to express their appreciation to Navin Khaneja for valuable discussions.
This work was supported by $\text{EPSRC, ERC, BBOI}$, QUCHIP(H2020-FETPROACT-3-2014), PICQUE(FP7-PEOPLE-2013-ITN), US Army Research Office(ARO) Grant W911NF-14-1-0133 and the Centre for Nanoscience and Quantum Information(NSQI). X.Z. acknowledges support from the National Key R \& D Program (Grant No. 2016YFA0301700), the National Young 1000 Talents Plan and Natural Science Foundation of Guangdong (2016A030312012). J.L.OB. acknowledges a Royal Society Wolfson Merit Award and a Royal Academy of Engineering Chair in Emerging Technologies. The experimental data are available for download from the Research Data Repository of University of Bristol at https://data.bristol.ac.uk/data/dataset/35xkv6pvafi8d23orogqgewm9u.

\section*{Appendix}
\textbf{Linear decomposition of a unitary operation.}
Here we show how to decompose a unitary quantum operation into the linear combination form. We first consider two-qubit unitary operations. By using the KAK decomposition~\cite{kraus2001optimal}, an arbitrary two-qubit unitary operation $U_{\text{SU(4)}}$ can be decomposed as 
\begin{align}
U_{\text{SU(4)}} = (U_1 \otimes V_1)U_D(U_2 \otimes V_2),
\end{align}
where $U_1$, $V_1$, $U_2$ and $V_2$ are single-qubit quantum gates, and $U_D$ is a non-factorable two-qubit gate responsible for the non-local characteristic of the gate $U$, which is given by
\begin{align}
U_D = \exp (-i(k_1\sigma _x\otimes \sigma _x + k_2\sigma _y\otimes \sigma _y +k_3\sigma _z\otimes \sigma _z)),
\label{eq:Ud}
\end{align}
where $k_i$ are real numbers, and $\sigma _x,$ $\sigma _y$ and $\sigma _z$ are Pauli matrices. 
Consider the facts that $\exp(iAx) = \cos (x)I + i\sin(x)A$ for an arbitrary real number $x$ and a matrix $A$ satisfying $A^2 = I$~\cite{nielsen2010quantum} and $\sigma _a \sigma _b = - \sigma _b \sigma _a = i \sigma _c$ for $\{a,b,c\}\in\{\{x,y,z\},\{y,z,x\},\{z,x,y\}\}$, we can obtain 
\begin{align}
&U_{\text{SU(4)}} \nonumber \\
&= (U_1 \otimes V_1) \cdot(\alpha_0 I\otimes I + \alpha_1 \sigma _x \otimes \sigma _x + \alpha_2 \sigma _y \otimes \sigma _y + \alpha_3 \sigma _z \otimes \sigma _z) \cdot(U_2 \otimes V_2) \nonumber \\
& = \alpha_0 U_1U_2 \otimes V_1V_2 + \alpha_1 U_1\sigma _xU_2 \otimes V_1\sigma _xV_2 + \alpha_2 U_1\sigma _yU_2 \otimes V_1\sigma _yV_2 + \alpha_3 U_1\sigma _zU_2 \otimes V_1\sigma _zV_2. 
\end{align}
where $\alpha_i$ ($i = 0, \cdots, 3$) are complex coefficients derived from $k_i$ ($i=1,2,3$) in Eq. \eqref{eq:Ud}. The details are shown in Supplementary Material, together with the explicit results of decomposing universal three-qubit unitaries. More generally, an arbitrary $n$-qubit quantum operation $U \in \text{SU}(2^n)$ can be decomposed as a linear combination of the tensor products of $n$ single qubit gates, by applying Cartan's KAK decomposition recursively~\cite{khaneja2001cartan}. {The computational complexity of applying Cartan's decomposition on a unitary $U \in \text{SU}(d)$ is $O(\mbox{poly}(d))$~\cite{khaneja2016time}, and thus it is not efficient for a general exponential-sized unitary. It is an open problem to find efficient ways for applying Cartan's decomposition on specific families of unitary, for example, multiple controlled-unitary operations.}

\noindent\textbf{Experimental setup.}
{The polarization-entangled photon pairs are generated by a spontaneous parametric down-conversion source using paired type-I BiBO crystal in sandwich configuration~\cite{rangarajan2009optimizing}, where a diagonally polarized, 120 mW, continuous-wave laser beam with central wavelength of 404 nm is focused at the centre of paired BiBO crystals with their optical axes orthogonally aligned to each other.} The generated photons pass through a PBS cube on the client's side and a PBS/BS (half-PBS, half-BS) cube on the server's side respectively, generating the spatially-entangled state
\begin{align}
( \left| {H_{1b}} \right\rangle \left| {H_{2b}} \right\rangle +\left| {V_{1r}} \right\rangle \left| {V_{2r}} \right\rangle) /{\sqrt{2}}.
\end{align}
The client can prepare an arbitrary polarization-state $\left| \phi \right\rangle$ by configuring $P_{1b}$ and $P_{1r}$---consisting of half- and quarter- waveplates and acting on spatial modes $1b$ and $1r$ respectively. The server configures the computational input state $\ket{\psi}$ for computation by $Q_{2b}$ and $Q_{2r}$ which act on the spatial modes $2b$ and $2r$ respectively.  Note here that we assume that the client informs the server of the computational input state $\ket{\psi}$ in advance. The two single-qubit gates $A$ and $B$ are configured by the server using two sets of wave plates, each consisting of quarter-, half- and quarter waveplates. When detecting two-photon coincidences between detectors at ports 1 and 2, the client implements the quantum computation $(\alpha A + \beta B) \left| \psi \right\rangle$ securely on the remote server.

{ 
\noindent\textbf{Comparison with related work.} Previous protocols in refs~\cite{Fisher2014,aharonov2008interactive,childs2005secure,dupuis2012actively,broadbent2013quantum} provide security by hiding the computation data from the server while the algorithm itself is exposed to the server. Blind quantum computing~\cite{arrighi2006blind,broadbent2009universal,barz2012demonstration} can hide all of the computation input, output and algorithm. Since our protocol focuses on hiding the computation algorithm, we present here a comparison with blind quantum computing as below:}
\begin{table}[h!]
\centering
\caption{Comparing our protocol with blind quantum computing.}
\label{my-label}
\footnotesize
\begin{tabular}{l |p{4.8cm} p{4.8cm}}
\hline
 & \textbf{Blind quantum computing} & \textbf{Our protocol} \\ 
 \hline
 Privacy & input, output and algorithm & algorithm \\
{Computation model} & measurement-based model & quantum circuit model \\
{Algorithm encoding} & consecutive adaptive single-qubit measurements & amplitudes of a quantum state \\
{Requirements for client} & perfect randomness source; creation of single-qubit states & creation of small-scale states \\
Requirements for server & generation of large cluster states & implementation of basic computation components \\
Communications & transmission of quantum states; classical measurement instructions & EPR channels; Bell measurement results\\
Universality & universal & limited number of linear combination terms \\
Feasibility & difficult & near-term implementation \\
\hline
\end{tabular}
\end{table}

\section*{Reference}
\providecommand{\newblock}{}


\clearpage
\pagebreak
\newpage

\title{Supplementary Material for: \\ Quantum processing by remote quantum control}

\setcounter{equation}{0}
\setcounter{figure}{0}
\setcounter{table}{0}
\setcounter{page}{1}
\setcounter{section}{0}
\renewcommand{\theequation}{S\arabic{equation}}
\renewcommand{\thefigure}{S\arabic{figure}}
\renewcommand{\thesection}{S\arabic{section}}
\normalsize

\section{Evolution for the proposed LCC}
Here we show the step-by-step evolution of the LCC described in main text. The $(n\times d)$-dimensional target subspace $T$ decomposes into $n$ $d$-dimensional subspaces, with the $j^{th}$ subspace spanned by basis elements ${\ket{jd}_T,\cdots, \ket{(j+1)d-1}_T}$. The $0^{th}$ subspace, spanned by the basis states $\left| 0 \right\rangle_T, \left| 1 \right\rangle_T, \cdots, \left| d-1 \right\rangle_T$ and encodes the computational input state, while all other subspaces have zero amplitudes. Therefore, the initial state for $T$ is of the form
\begin{align}
\left| \Psi_{ext}  \right\rangle_T  = {\sum\nolimits_{j = 0}^{d - 1} {{\beta _j}{\left| j \right\rangle}_T} } + {\sum\nolimits_{j = d}^{nd - 1} {{0}{\left| j \right\rangle}_T} },
\end{align}
where $d$ represents the dimension for the target computation, $k$ represents the number of control qubits, and $n=2^k$ (as defined in main text).

We define ${\left| \Psi_{ext} \right\rangle_T ^s}$ ($s = 0, 1, \cdots, n-1$) as
\begin{align}
\left| \Psi_{ext}  \right\rangle_T ^s  = {\sum\nolimits_{j = 0}^{sd - 1} {0{\left| j \right\rangle}_T} } + {\sum\nolimits_{j =sd }^{(s+1)d - 1} {{\beta _j}{\left| j \right\rangle}_T} } + {\sum\nolimits_{j = (s+1)d}^{nd - 1} {{0}{\left| j \right\rangle}_T} }
\end{align}
where only the basis of the $s^{th}$ subspace have non-zero amplitudes. The initial state ${\left| \Psi_{ext}  \right\rangle}_T$ can then be represented as ${\left| \Psi_{ext}  \right\rangle_T ^0}$. $X^{(0,j)}$ exchanges corresponding basis elements between $0^{th}$ and $j^{th}$ subspaces, which equivalently swaps the two states $\left| \Psi_{ext}  \right\rangle_T ^0$ and $\left| \Psi_{ext}  \right\rangle_T ^j$.
The sum operation $V_{sum} =  \oplus _{j = 0}^{n - 1}{V_j^{(j)}}$ is an $n \times d$ dimension quantum operation, where $V_j^{(j)}$ implements the $d$-dimension quantum operation $V_j$ on $j^{th}$ subspace of $T$.

The $k$-qubit control $\ket{\phi}_C$ can be expanded as follows (note $n=2^k$),
\begin{align}
\left| \phi  \right\rangle_C  ={\sum\nolimits_{j = 0}^{n- 1} {\alpha _j{\left| j \right\rangle}} } = {\alpha _0}\overbrace {\left| {00 \cdots 0} \right\rangle }^k + {\alpha _1}\overbrace {\left| {00 \cdots 1} \right\rangle }^k +  \cdots  + {\alpha _{n - 1}}\overbrace {\left| {11 \cdots 1} \right\rangle }^k.
\end{align}
The evolution of the LCC can be obtained as follows, with time going from left to right:
\begin{align}
&\left| \phi  \right\rangle_C \left| \Psi_{ext}  \right\rangle_T^0 \nonumber \\
&= {\alpha _0}\overbrace {\left| {00 \cdots 0} \right\rangle }^k{\left| \Psi_{ext}  \right\rangle_T^0} + {\alpha _1}\overbrace {\left| {00 \cdots 1} \right\rangle }^k{\left| \Psi_{ext}  \right\rangle_T^0} +  \cdots  + {\alpha _{n - 1}}\overbrace {\left| {11 \cdots 1} \right\rangle }^k{\left| \Psi_{ext}  \right\rangle_T^0}\\
 &\to {\alpha _0}\overbrace {\left| {00 \cdots 0} \right\rangle }^k{\left| \Psi_{ext}  \right\rangle_T^0} + {\alpha _1}\overbrace {\left| {00 \cdots 1} \right\rangle }^k{\left| \Psi_{ext}  \right\rangle_T^1} +  \cdots  + {\alpha _{n - 1}}\overbrace {\left| {11 \cdots 1} \right\rangle }^k{\left| \Psi_{ext}  \right\rangle_T^{n-1}}\\
 &\to {\alpha _0}\overbrace {\left| {00 \cdots 0} \right\rangle }^kV_{sum}{\left| \Psi_{ext}  \right\rangle_T^0} + {\alpha _1}\overbrace {\left| {00 \cdots 1} \right\rangle }^kV_{sum}{\left| \Psi_{ext}  \right\rangle_T^1} +  \cdots  + {\alpha _{n - 1}}\overbrace {\left| {11 \cdots 1} \right\rangle }^kV_{sum}{\left| \Psi_{ext}  \right\rangle_T^{n-1}}\\
 &= {\alpha _0}\overbrace {\left| {00 \cdots 0} \right\rangle }^k{\left| \Psi_{ext}  \right\rangle_T ^{{V_0^{(0)}},0}} + {\alpha _1}\overbrace {\left| {00 \cdots 1} \right\rangle }^k{\left| \Psi_{ext}  \right\rangle_T ^{{V_1^{(1)}},1}} +  \cdots  + {\alpha _{n - 1}}\overbrace {\left| {11 \cdots 1} \right\rangle }^k{\left| \Psi_{ext}  \right\rangle_T ^{{V_{n - 1}^{(n-1)}},n - 1}}\\
 &\to {\alpha _0}\overbrace {\left| {00 \cdots 0} \right\rangle }^k{\left| \Psi_{ext}  \right\rangle_T ^{{V_0^{(0)}},0}} + {\alpha _1}\overbrace {\left| {00 \cdots 1} \right\rangle }^k{\left| \Psi_{ext}  \right\rangle_T ^{{V_1^{(0)}},0}} +  \cdots  + {\alpha _{n - 1}}\overbrace {\left| {11 \cdots 1} \right\rangle }^k{\left| \Psi_{ext}  \right\rangle_T ^{{V_{n - 1}^{(0)}},0}}\\
 &\to \frac{1}{{{2^{k/2}}}}\overbrace {\left| {00 \cdots 0} \right\rangle }^k\left( {{\alpha _0}{\left| \Psi_{ext}  \right\rangle_T^{{V_0^{(0)}},0}} + {\alpha _1}{\left| \Psi_{ext}  \right\rangle_T^{{V_1^{(0)}},0}} +  \cdots  + {\alpha _{n - 1}}{\left| \Psi_{ext}  \right\rangle_T^{{V_{n - 1}^{(0)}},0}}} \right) + \nonumber \\
 &~~~~~ \frac{1}{{{2^{k/2}}}}\overbrace {\left| {00 \cdots 1} \right\rangle }^k\left( {{\alpha _0}{\left| \Psi_{ext}  \right\rangle_T^{{V_0^{(0)}},0}} - {\alpha _1}{\left| \Psi_{ext}  \right\rangle_T^{{V_1^{(0)}},0}} +  \cdots  - {\alpha _{n - 1}}{\left| \Psi_{ext}  \right\rangle_T^{{V_{n - 1}^{(0)}},0}}} \right) + \cdots + \nonumber \\
 &~~~~~ \frac{1}{{{2^{k/2}}}}\overbrace {\left| {11 \cdots 1} \right\rangle }^k\left( {{\alpha _0}{\left| \Psi_{ext}  \right\rangle_T^{{V_0^{(0)}},0}} - {\alpha _1}{\left| \Psi_{ext}  \right\rangle_T^{{V_1^{(0)}},0}} +  \cdots  + {{\left( { - 1} \right)}^k}{\alpha _{n - 1}}{\left| \Psi_{ext}  \right\rangle_T^{{V_{n - 1}^{(0)}},0}}} \right)
\end{align}
Note here that ${{\left| \Psi_{ext}  \right\rangle_T^{{V_j^{(k)}},k}}}$ $(j,k = 0, 1, \cdots , n-1)$ means that the $d$-dimension operation $V_j$ acts on the $k^{th}$ subspace of $T$ where $T$ has the state of $\left| \Psi_{ext}  \right\rangle_T^k$.

When the $k$ control qubits are all measured to be 0 in the computational basis, the resulting state of $T$ is obtained as
\begin{align}
&{{\alpha _0}{{\left| \Psi_{ext}  \right\rangle }_T^{{V_0^{(0)}},0}} + {\alpha _1}{{\left| \Psi_{ext}  \right\rangle }_T^{{V_1^{(0)}},0}} +  \cdots  + {\alpha _{n - 1}}{{\left| \Psi_{ext}  \right\rangle }_T^{{V_{n - 1}^{(0)}},0}}} \nonumber \\
=& {\left({\sum\nolimits_{j = 0}^{n - 1} {\alpha _j V_j^{(0)} }}\right)}{\sum\nolimits_{j = 0}^{d - 1} {\beta_j{\left| j \right\rangle_T}} } + {\sum\nolimits_{j = d}^{nd - 1} {0{\left| j \right\rangle_T}} }.
\end{align}
This shows that the operation $U = \sum\nolimits_{j = 0}^{n - 1} {\alpha _j V_j } $ is implemented on the state $\left| \psi  \right\rangle  = {\sum\nolimits_{j = 0}^{d - 1} {\beta _j{\left| j \right\rangle_T}} }$ which lies in the $0^{th}$ subspace of $T$. The success probability of this LCC is $(\frac{1}{{{2^{k/2}}}})^2 = \frac{1}{n}$, decreasing polynomially with the length of the gate sequence for operations being combined.

\section{Linear decomposition of unitary quantum operation} 
Here we present more details of the linear decomposition of a unitary quantum operation. We start by showing the explicit linear decomposition of universal two-qubit quantum operation. It has been shown that an arbitrary two-qubit operation $U_{\text{SU(4)}} \in \text{SU(4)}$ can be decomposed as~\cite{kraus2001optimal}:
\begin{align}
U_{\text{SU(4)}} = (U_1 \otimes V_1)U_D(U_2 \otimes V_2), \label{eq:2qubitKAK}
\end{align}
where $U_1$, $V_1$, $U_2$ and $V_2$ are single-qubit quantum gates, and $U_D$ is a non-factorable two-qubit gate responsible for the non-local characteristic of the gate $U$, which is given by
\begin{align}
U_D = \exp (-i(k_1\sigma _x\otimes \sigma _x + k_2\sigma _y\otimes \sigma _y +k_3\sigma _z\otimes \sigma _z)),
\end{align}
where $k_i$ are real numbers, and $\sigma _x,$ $\sigma _y$ and $\sigma _z$ are Pauli matrices. Define a matrix $M$ as
\begin{align}
M = \frac{1}{\sqrt{2}}\begin{pmatrix}
1 & 0 & 0 & i \\
0 & i & 1 & 0 \\
0 & i & -1 & 0 \\
1 & 0 & 0 & -i \\
\end{pmatrix},
\end{align}
and then $U_1 \otimes V_1$ and $U_2 \otimes V_2$ can be obtained as
\begin{align}
U_1\otimes V_1 & = MLM^\dag \label{eq:u1v1} \\
U_2\otimes V_2 & = MRM^\dag  \label{eq:u2v2}
\end{align}
where $L$ and $R$ are two real orthogonal matrices that are obtained by performing the simultaneous singular value decomposition for $U'_R = \text{Real}(M^{\dag} U_{\text{SU(4)}}M)$ (real part) and $U'_I = \text{Imag}({M^{\dag} U_{\text{SU(4)}}M})$ (imaginary part), together with two non-negatively real diagonal matrices $D_R$ and $D_I$. They satisfy that
\begin{align}
D_R = L^\dag U'_R R, \\
D_I = L^\dag U'_I R.
\end{align}
$U_D$ and further $k_i$'s can be obtained through 
\begin{align}
U_D = M(D_R+iD_I)M^\dag .
\end{align}  
A step-by-step procedure for obtaining the decomposition result in Eq. \eqref{eq:2qubitKAK} is given in ref~\cite{tucci2005introduction}.

Consider the facts that 
\begin{align}
\exp(iAx) = \cos (x)I + i\sin(x)A
\end{align}
where $x$ is an arbitrary real number and $A$ is a matrix satisfying $A^2 = I$~\cite{nielsen2010quantum} and
\begin{align}
&\sigma _x \sigma _y = - \sigma _y \sigma _x = i \sigma _z, \\
& \sigma _y \sigma _z = - \sigma _z \sigma _y = i \sigma _x, \\
&\sigma _z \sigma _x = - \sigma _x \sigma _z = i \sigma _y.
\end{align}
$U_{\text{SU}(4)}$ can be rewritten into the following form:
\begin{align}
U_{\text{SU(4)}} &= (U_1 \otimes V_1) \cdot(\alpha_0 I\otimes I + \alpha_1 \sigma _x \otimes \sigma _x + \alpha_2 \sigma _y \otimes \sigma _y + \alpha_3 \sigma _z \otimes \sigma _z) \cdot(U_2 \otimes V_2) \nonumber \\
& = \alpha_0 U_1IU_2 \otimes V_1IV_2 + \alpha_1 U_1\sigma _xU_2 \otimes V_1\sigma _xV_2 + \alpha_2 U_1\sigma _yU_2 \otimes V_1\sigma _yV_2 + \alpha_3 U_1\sigma _zU_2 \otimes V_1\sigma _zV_2.
\end{align}
where $\alpha_0$, $\alpha_1$, $\alpha_2$ and $\alpha_3$ are complex coefficients defined as
\begin{align}
\alpha_0& = (\cos(k_1)\cos(k_2)\cos(k_3) - i\sin(k_1)\sin(k_2)\sin(k_3)),  \nonumber \\
\alpha_1&=( \cos(k_1)\sin(k_2)\sin(k_3) - i \sin(k_1)\cos(k_2)\cos(k_3)), \nonumber \\
\alpha_2&=(\sin(k_1)\cos(k_2)\sin(k_3) - i \cos(k_1)\sin(k_2)\cos(k_3) ),  \nonumber \\
\alpha_3&=(\sin(k_1)\sin(k_2)\cos(k_3) - i\cos(k_1)\cos(k_2)\sin(k_3) ). 
\end{align}
This shows that an arbitrary two-qubit operation can be decomposed into a linear combiantion of four terms, each of which is a tensor product of two single-qubit quantum gates. Similarly, an arbitrary three-qubit quantum operation $U_{\text{SU(8)}} \in \text{SU(8)}$ can be decomposed as~\cite{vatan2004realization}:
\begin{align}
\label{eq:SU8}
U_{\text{SU(8)}}= (A_4 \otimes B_4)N_2(A_3\otimes B_3)M(A_2 \otimes B_2)N_1(A_1 \otimes B_1),
\end{align}
where $A_i$ is two-qubit gate, $B_i$ is single-qubit gate, $N_1$, $N_2$ and $M$ are defined as  
\begin{align}
N_k &= \exp(i(\alpha_{0}^{(k)}\sigma _x \otimes \sigma _x \otimes \sigma _z + \alpha_1^{(k)}\sigma _y \otimes \sigma _y \otimes \sigma _z +\alpha_2^{(k)}\sigma _z \otimes \sigma _z \otimes \sigma _z)) \label{eq:Nk}\\
M &= \exp(i(\beta_0\sigma _x \otimes \sigma _x \otimes \sigma _x + \beta_1\sigma _y \otimes \sigma _y \otimes \sigma _x + \beta_2\sigma _z \otimes \sigma _z \otimes \sigma _x + \beta_3I \otimes I \otimes \sigma _x)).\label{eq:M}
\end{align}
Here $\alpha_i^{(k)}$ and $\beta_j$ are real numbers. Applying similar algebra as that used in the case of two-qubit operations, we can obtain the linear-combination decomposition form of $U_{\text{SU(8)}}$ where each of term is a tensor-product of three single-qubit gates. 

More generally, an arbitrary $n$-qubit quantum operation $U \in \text{SU}(2^n)$ can be decomposed as 
\begin{align}
U = K_1AK_2,
\end{align}
where $K_1, K_2 \in \text{SU}(2^{n-1})\otimes \text{SU}(2^{n-1})\otimes U(1)$ and $A \in \exp(h)$, with $h$ being a Cartan subalgebra of the Riemannian symmetric space $\text{SU}(2^n)/\text{SU}(2^{n-1})\otimes \text{SU}(2^{n-1}) \otimes U(1)$~\cite{khaneja2001cartan}. A recursive formula can then be obtained by further decomposing $K_1$ and $K_2$ in terms of the elements of $\text{SU}(2^{n-2})\otimes \text{SU}(2^{n-2})\otimes U(1)$ and so on~\cite{khaneja2001cartan}. Finally, we can rewrite the given $n$-qubit operation into a linear combination of tensor products of $n$ single-qubit gates. It is easy to find that such a linear-combination decomposition is not efficient---it generally requires exponentially many linear terms. 

However, in some cases, the number of the linear terms for the decomposition of a given operation is much less. We have mentioned that in the main text an arbitrary controlled-unitary operation can be rewritten into the linear combination of four terms. Here is another example: when the coefficients $\alpha_i^{(k)}$ ($i=0,1,2; k=1,2$), $\beta_1$, $\beta_2$ and $\beta_3$ in Eq.~\eqref{eq:Nk} and \eqref{eq:M} are all zeros, the corresponding linear decomposition of $U_{\text{SU(8)}}$ will include only two terms as follows:
\begin{align}
U_\text{SU(8)} &= (A_4 A_3 \otimes B_4 B_3)\exp(i\beta_0\sigma_x^{\otimes 3})(A_2A_1 \otimes B_2B_1) \nonumber \\
& = \cos(\beta_0)(A_4 A_3A_2 A_1)\otimes(B_4B_3B_2B_1) + i\sin(\beta_0)(A_4 A_3\sigma_x^{\otimes 2}A_2 A_1)\otimes(B_4B_3\sigma_xB_2B_1)
\end{align}
where $A_i$ and $B_i$ ($i = 1,\cdots, 4$) are defined as in Eq.~\eqref{eq:SU8} and $\sigma_x^{\otimes 3}=\sigma_x \otimes \sigma_x \otimes \sigma_x $. 

\section{Security analysis of the proposed protocol}
\begin{figure}[tbp]
\centering
\includegraphics[scale =1]{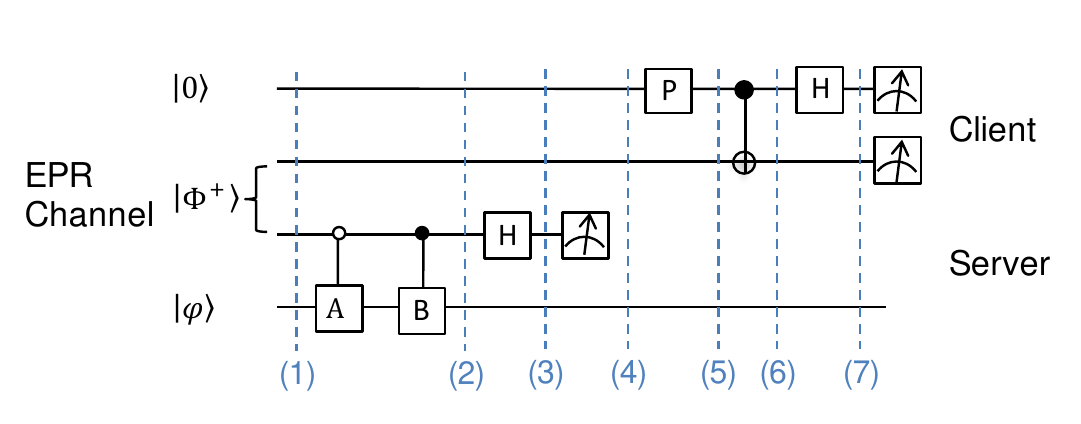}
\caption{ \textbf{Linear-combining two known operations by remote one-qubit control.} $\ket{\Phi^+} = \frac{1}{\sqrt{2}}(\ket{00}+\ket{11})$ is the EPR state used for quantum teleportation. $\ket{\varphi}$ is a quantum regiseter state used for the input of quantum computation, and $A$ and $B$ are two corresponding quantum operations with the same size. $P$ represents the single-qubit operation to configure the one-qubit control state: $\alpha \ket{0} + \beta\ket{1}$.
}
\label{security}
\end{figure}

The security of our proposed protocol has been discussed in the main text. Here we present more details of the security analysis for one-qubit control quantum processing (see Figure 3 in the main text): we have chosen the case where the client only sends a one-qubit control state to the server to linearly combine two quantum operations $A$ and $B$. We also assume that $A$ and $B$ are not black-box operations to the server, and thus the server can implement the linear-combination operation using the circuit shown in Fig.1(A) in main text. This assumption does not weaken our security arguments, since in our protocol the privacy is kept just through hiding the linear coefficients. We assume the server runs the LCC before the client teleports the control state. The corresponding circuit is shown in Fig.~\ref{security}, with the step-by-step evolution states being labeled. The evolution of the circuit is then given as follows. 

\begin{align}
&(1):~\frac{1}{\sqrt{2}}(\ket{0}_1\ket{0}_2\ket{0}_3\ket{\varphi}+ \ket{0}_1\ket{1}_2\ket{1}_3\ket{\varphi}) \\
&(2):~\frac{1}{\sqrt{2}}(\ket{0}_1\ket{0}_2\ket{0}_3A\ket{\varphi}+ \ket{0}_1\ket{1}_2\ket{1}_3B\ket{\varphi}) \\
&(3):~\frac{1}{2}(\ket{0}_1\ket{0}_2\ket{0}_3A\ket{\varphi} +\ket{0}_1\ket{0}_2\ket{1}_3A\ket{\varphi}+ \ket{0}_1\ket{1}_2\ket{0}_3B- \ket{0}_1\ket{0}_2\ket{1}_3B\ket{\varphi}) \\
&(4):~\frac{1}{\sqrt{2}}(\ket{0}_1\ket{0}_2A\ket{\varphi}+\ket{0}_1\ket{1}_2B\ket{\varphi})\\
&(5):~\frac{1}{\sqrt{2}}(\alpha\ket{0}_1\ket{0}_2A\ket{\varphi}+\beta\ket{1}_1\ket{0}_2A\ket{\varphi}+\alpha\ket{0}_1\ket{1}_2B\ket{\varphi}+\beta\ket{1}_1\ket{1}_2B\ket{\varphi}) \\
&(6):~\frac{1}{\sqrt{2}}(\alpha\ket{0}_1\ket{0}_2A\ket{\varphi}+\beta\ket{1}_1\ket{1}_2A\ket{\varphi}+\alpha\ket{0}_1\ket{1}_2B\ket{\varphi}+\beta\ket{1}_1\ket{0}_2B\ket{\varphi}) \\
&(7):~\frac{1}{2}(\alpha\ket{0}_1\ket{0}_2A\ket{\varphi} + \alpha\ket{1}_1\ket{0}_2A\ket{\varphi}
+\beta\ket{0}_1\ket{1}_2A\ket{\varphi} - \beta\ket{1}_1\ket{1}_2A\ket{\varphi} \nonumber \\
&~~~~~~~~~~+\alpha\ket{0}_1\ket{1}_2B\ket{\varphi} + \alpha\ket{1}_1\ket{1}_2B\ket{\varphi}
+\beta\ket{0}_1\ket{0}_2B\ket{\varphi} - \beta\ket{1}_1\ket{0}_2B\ket{\varphi})
\end{align}
Here, the subscripts ``1'', ``2'' and ``3'' represent the client's local qubit and the EPR qubits owned by the client and the server respectively, the same below. When the client measures the qubit 1 and qubit 2 to be ``0'' in the computational basis, the state of the quantum register ($\ket{\varphi}$) will be $(\alpha A + \beta B)\ket{\varphi}$. In this case, the server measures the control qubit before the client prepares it, and thus the linear coefficients are kept hidden from the server.

\begin{figure}[tbp]
\centering
\includegraphics[scale=1]{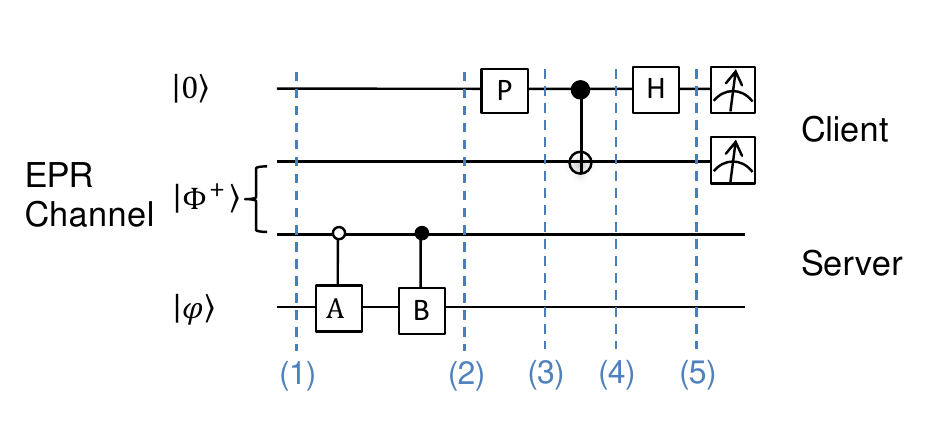}
\caption{
\textbf{Server lies in the process of linear-combining two known operations by remote one-qubit control.} $\ket{\Phi^+} = \frac{1}{\sqrt{2}}(\ket{00}+\ket{11})$ is the EPR state used for quantum teleportation. $\ket{\varphi}$ is a quantum regiseter state used for the input of quantum computation, and $A$ and $B$ are two corresponding quantum operations with the same size. $P$ represents the single-qubit operation to configure the one-qubit control state: $\alpha \ket{0} + \beta\ket{1}$.
}
\label{serverlies}
\end{figure}

Next, we consider the case where the server lies to the client that he had measured the qubit 3 but actually he did not. The corresponding circuit is shown in Fig.~\ref{serverlies}, with step-by-step evolution state being labeled. The evolution of this circuit is then given as follows.

\begin{align}
&(1):~\frac{1}{\sqrt{2}}(\ket{0}_1\ket{0}_2\ket{0}_3\ket{\varphi}+ \ket{0}_1\ket{1}_2\ket{1}_3\ket{\varphi}) \\
&(2):~\frac{1}{\sqrt{2}}(\ket{0}_1\ket{0}_2\ket{0}_3A\ket{\varphi}+ \ket{0}_1\ket{1}_2\ket{1}_3B\ket{\varphi}) \\
&(3):~\frac{1}{\sqrt{2}}(\alpha\ket{0}_1\ket{0}_2\ket{0}_3A\ket{\varphi}+\beta\ket{1}_1\ket{0}_2\ket{0}_3A\ket{\varphi}+ \alpha\ket{0}_1\ket{1}_2\ket{1}_3B\ket{\varphi}+\beta\ket{1}_1\ket{1}_2\ket{1}_3B\ket{\varphi}) \\
&(4): ~\frac{1}{\sqrt{2}}(\alpha\ket{0}_1\ket{0}_2\ket{0}_3A\ket{\varphi}+\beta\ket{1}_1\ket{1}_2\ket{0}_3A\ket{\varphi}+ \alpha\ket{0}_1\ket{1}_2\ket{1}_3B\ket{\varphi}+\beta\ket{1}_1\ket{0}_2\ket{1}_3B\ket{\varphi}) \\
&(5):~\frac{1}{2}(\alpha\ket{0}_1\ket{0}_2\ket{0}_3A\ket{\varphi}+\alpha\ket{1}_1\ket{0}_2\ket{0}_3A\ket{\varphi}+\beta\ket{0}_1\ket{1}_2\ket{0}_3A\ket{\varphi}-\beta\ket{1}_1\ket{1}_2\ket{0}_3A\ket{\varphi}\nonumber \\
&~~~~~~~+ \alpha\ket{0}_1\ket{1}_2\ket{1}_3B\ket{\varphi}+\alpha \ket{1}_1\ket{1}_2\ket{1}_3B\ket{\varphi}+\beta\ket{0}_1\ket{0}_2\ket{1}_3B\ket{\varphi}-\beta \ket{1}_1\ket{0}_2\ket{1}_3B\ket{\varphi})
\end{align}
When the client measures the qubit 1 and qubit 2 to be ``0'' in the computational basis, the state of remaining qubits will be
\begin{align}
\ket{\Psi} =\alpha \ket{0}_3A\ket{\varphi} + \beta \ket{1}_3 B \ket{\varphi}.
\end{align}
Now we need to know if the server can extract the information of the control state without being found by the client. This requires that the server can extract the control state $\ket{\phi}_C = \alpha \ket{0} + \beta \ket{1}$ while the client obtains the correct result $(\alpha A + \beta B)\ket{\varphi}$. The server can only achieve this if there exists a quantum operation $U_s$ that satisfies
\begin{align}
(\alpha \ket{0}+ \beta \ket{1})(\alpha A + \beta B)\ket{\varphi} = U_s (\alpha \ket{0}A\ket{\varphi} + \beta \ket{1} B \ket{\varphi}).
\end{align}
Such an $U_s$ does not exist for unknown $\alpha$ and $\beta$, since the no-cloning theorem forbids faithful copying of unknown quantum states.

We can see this more clearly from an explicit example. Suppose $A = I$, $B=X$ and $\ket{\varphi} = \ket{0}$, then the state that the server expected is 
\begin{align}
\ket{\Phi}_{exp} & =(\alpha \ket{0}+ \beta \ket{1})(\alpha A + \beta B)\ket{\varphi}= \begin{pmatrix} \alpha \\ \beta \end{pmatrix}
\otimes \left( \begin{pmatrix} \alpha & \beta \\ \beta & \alpha \end{pmatrix}\cdot \begin{pmatrix} 1 \\ 0 \end{pmatrix} \right) =\begin{pmatrix} \alpha^2 \\ \alpha \beta \\ \beta\alpha \\ \beta^2 \end{pmatrix}
\label{eq:serverExp}
\end{align} 
The state $\ket{\Psi}$ will be
\begin{align}
\ket{\Psi} & = \alpha \ket{0}_3I\ket{0} + \beta \ket{1}_3 X \ket{0} = \alpha \ket{0}_3\ket{0} + \beta \ket{1}_3 \ket{1}  = \begin{pmatrix} \alpha \\ 0 \\ 0 \\ \beta \end{pmatrix} 
\label{eq:serverOb}
\end{align}
Comparing Eqs. \eqref{eq:serverExp} and \eqref{eq:serverOb}, there is no $U_s$ that satisfies $\ket{\Phi}_{exp} = U_s \ket{\Psi}$ for general $\alpha$ and $\beta$.


\section{Linear decomposition of single-qubit gate}
An arbitrary single-qubit quantum operation $U_{\text{SU(2)}} \in \text{SU(2)}$ can be written into the form~\cite{khaneja2000cartan,chatzisavvas2009improving} 
\begin{align}
U_{\text{SU(2)}}  = \exp(-i(d_1\sigma_x + d_2 \sigma_y + d_3 \sigma_z))
\end{align}
where $d_i$ ($i=1,2,3$) is real number. We can rewrite $U_{\text{SU(2)}}$ in the linear-combination form as follows
\begin{align}
U_{\text{SU(2)}} &= (\cos(d_1)I - i\sin(d_1)\sigma_x)(\cos(d_2)I - i\sin(d_2)\sigma_y)(\cos(d_3)I - i\sin(d_3)\sigma_z) \\
& = \alpha_0 I + \alpha_1 \sigma_x + \alpha_2 \sigma_y + \alpha_3 \sigma_z \nonumber
\end{align}
where $\alpha_0$, $\alpha_1$, $\alpha_2$ and $\alpha_3$ are given by
\begin{align}
\alpha_0 &= \cos(d_1)\cos(d_2)\cos(d_3) - \sin(d_1)\sin(d_2)\sin(d_3), \\
\alpha_1 &= -  i(\cos(d_1)\sin(d_2)\sin(d_3) + \sin(d_1)\cos(d_2)\cos(d_3)), \\
\alpha_2 &=  -  i(\cos(d_1)\sin(d_2)\cos(d_3) - \sin(d_1)\cos(d_2)\sin(d_3)),\\
\alpha_3 &= -  i(\cos(d_1)\cos(d_2)\sin(d_3) + \sin(d_1)\sin(d_2)\cos(d_3)).
\end{align}
i.e., an arbitrary single-qubit unitary operation can be decomposed as a linear combination of four terms: the identity and three Pauli matrices.

\section{Further experimental results}
\begin{figure*}[h]
\includegraphics[width=1\textwidth]{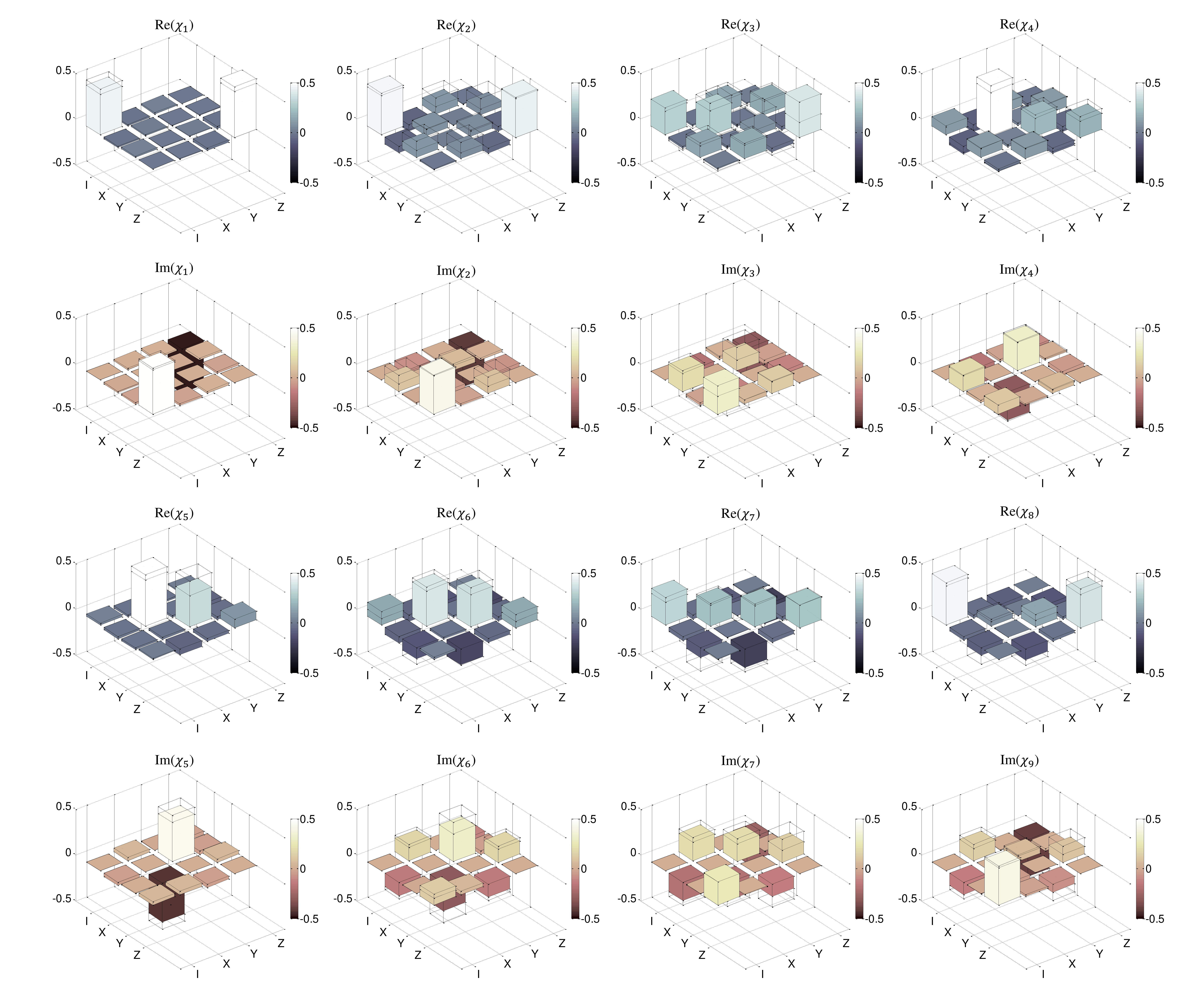}
\caption{
{\textbf{Experimental reconstructed $\chi$ matrices with ideal theoretical predictions overlaid:}
A single half-waveplate enables preparation of arbitrary one-qubit quantum control state $\ket{\phi}$ with real amplitudes. By setting half-waveplate angle to be $0^{\circ}$,  $11.25^{\circ}$, $22.5^{\circ}$, $33.75^{\circ}$, $45^{\circ}$, $56.25^{\circ}$, $67.5^{\circ}$ and $78.75^{\circ}$, $\left| \phi \right\rangle$ will be the state $\left| {0} \right\rangle$, $0.9239\left| {0} \right\rangle + 0.3827\left| {1} \right\rangle$, $0.7071\left| {0} \right\rangle + 0.7071\left| {1} \right\rangle$, $0.3827\left| {0} \right\rangle + 0.9239\left| {1} \right\rangle$, $\left| {1} \right\rangle$, $-0.3827\left| {0} \right\rangle + 0.9239\left| {1} \right\rangle$, $-0.7071\left| {0} \right\rangle + 0.7071\left| {1} \right\rangle$ and $-0.9239\left| {0} \right\rangle + 0.3827\left| {1} \right\rangle$. The eight corresponding constructed operations are denoted as $U_1$, $U_2$, $U_3$, $U_4$, $U_5$, $U_6$, $U_7$ and $U_8$ respectively. The maximum-likelihood technique was used to reconstruct the $\chi$ matrices from the experimental data. The matrix $\chi_i$($i=1,\cdots,8$) corresponds to the operation $U_i$($i=1,\cdots,8$). Both the real and imaginary part of each matrix are shown, with their ideal theoretical values overlaid.
The achieved process fidelities are $99.16\pm0.37\%$, $94.38\pm 0.87\%$, $94.79\pm0.85\%$, $91.94\pm1.10\%$, $95.99\pm 1.08\%$, $95.98\pm0.73\%$, $95.65\pm 0.91\%$ and $96.94\pm0.62\%$ respectively.
}}
\label{results5-8}
\end{figure*}

When the two single-qubit gates $A$ and $B$ are set to be
\begin{align}
A = \left( {\begin{array}{*{20}{c}}
\frac{1-i}{{\sqrt 2 }}&{ 0}\\
{ 0}&\frac{-1-i}{{\sqrt 2 }}
\end{array}} \right),
B = \left( {\begin{array}{*{20}{c}}
{0}&\frac{1+i}{{\sqrt 2 }}\\
\frac{1-i}{{\sqrt 2 }}&{0}
\end{array}} \right),
\end{align}
the client can always implement unitary operation $U = \alpha A + \beta B$ by teleporting an arbitrary one-qubit control state $\ket{\phi}=\alpha \left| {{0}} \right\rangle  + \beta \left| {1} \right\rangle $ with $\alpha$ and $\beta$ being real numbers.
By just using a single half-waveplate in $P$, the polarization state of the photon on the client's side, i.e., $\ket{\phi}$, can be configured into any single-qubit state with real amplitudes.  We set the angle of half-waveplate into $0^{\circ}$, $11.25^{\circ}$, $22.5^{\circ}$, $45^{\circ}$, $56.25^{\circ}$, $67.5^{\circ}$, $78.75^{\circ}$, and thus, eight different unitary operations denoted as $U_i$($i=1,2,\cdots, 8$) are implemented by the client. We performed quantum process tomography for each operation and reconstructed their process matrices from experimental data using maximum-likelihood estimation technique. The reconstructed process matrices are shown in Fig.~\ref{results5-8}, with corresponding process fidelities. The errors are estimated by adding random noise to the raw date obtained experimentally assuming Poissonian statistics, and then performing the reconstructions many times.  

\begin{figure*}[h]
\includegraphics[width=1\textwidth]{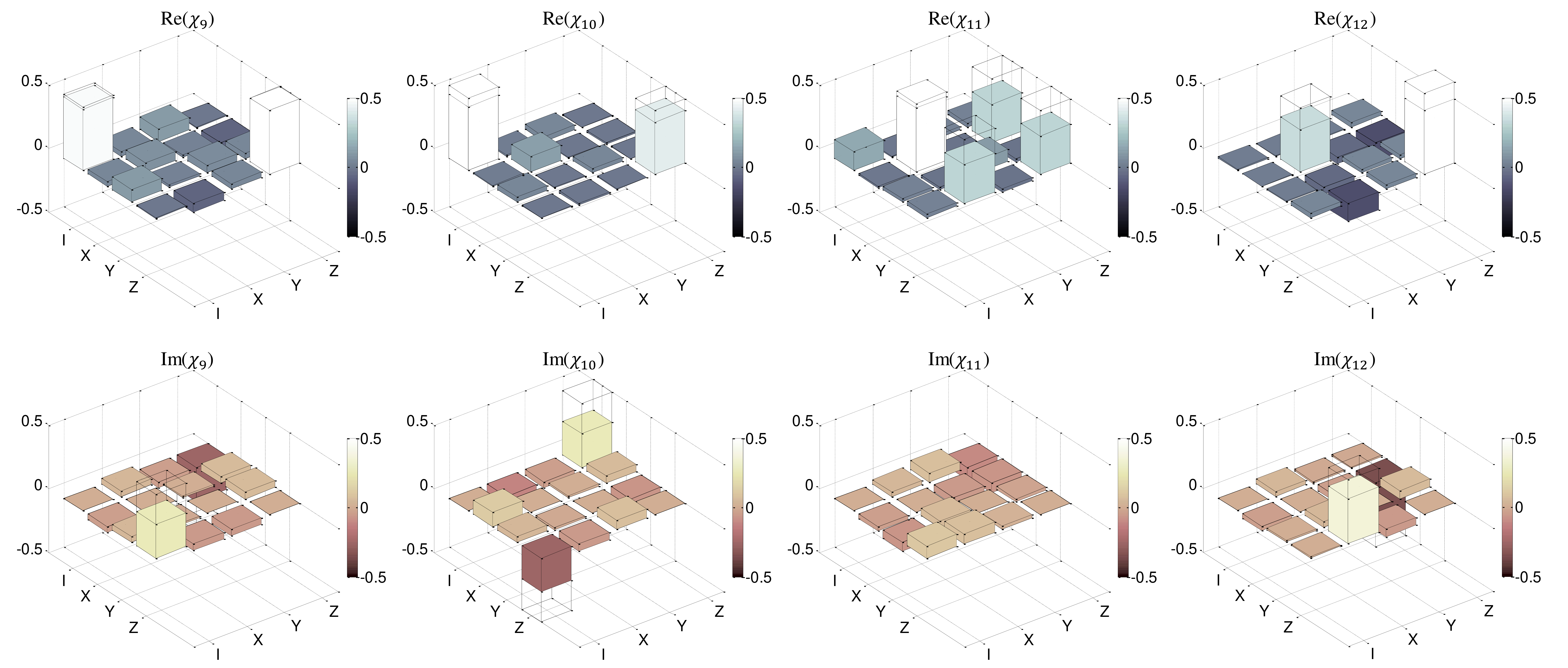}
\caption{
{\textbf{Experimental reconstructed $\chi$ matrices with ideal theoretical predictions overlaid:}
Four quantum operations $U_9=\frac{1}{{\sqrt 2 }}I + \frac{i}{{\sqrt 2 }}Z$, $U_{10}=\frac{1}{{\sqrt 2 }}I - \frac{i}{{\sqrt 2 }}Z$, $U_{11}=\frac{1}{{\sqrt 2 }}X + \frac{1}{{\sqrt 2 }}Z$ and $U_{12}=\frac{1}{{\sqrt 2 }}X + \frac{i}{{\sqrt 2 }}Z$ are implemented. Here $U_9$, $U_{10}$ and $U_{11}$ are unitary operations. $U_{12}$ is a non-unitary operation, which can filter out $\left| {L} \right\rangle ( =(\left| {0} \right\rangle - i\left| {1} \right\rangle)/{{\sqrt 2 }}  )$ and project all other basis state onto $\left| {{L}} \right\rangle $. The maximum-likelihood technique was used to reconstruct the $\chi$ matrices from the experimental data. The matrix $\chi_i$ ($i=9,\cdots,12$) corresponds to the operation $U_i$ ($i=9,\cdots,12$).
The obtained fidelities are 91.05$\pm$1.51$\%$, 90.16$\pm$1.91$\%$, 91.67$\pm$0.62$\%$ and 88.56$\pm$1.58$\%$ respectively.
}}
\label{results_more}
\end{figure*}

We also tested other configurations of $A$ and $B$: $A = I$ (Identity), $B = Z$ (Pauli-Z) and $A = X$ (Pauli-X), $B = Z$. By transmitting different one-qubit control state $\ket{\phi}$, the client implements various quantum operations on the server's side as follows:
\begin{align}
U_9=\frac{1}{{\sqrt 2 }}I + \frac{i}{{\sqrt 2 }}Z,~ U_{10}=\frac{1}{{\sqrt 2 }}I - \frac{i}{{\sqrt 2 }}Z, ~U_{11}=\frac{1}{{\sqrt 2 }}X + \frac{1}{{\sqrt 2 }}Z,~U_{12}=\frac{1}{{\sqrt 2 }}X + \frac{i}{{\sqrt 2 }}Z.
\end{align}
The reconstructed process matrices for these operations are shown in Fig.~\ref{results_more}, with corresponding process fidelities. The errors are estimated in the same way as mentioned above.

\providecommand{\newblock}{}


\end{document}